\documentclass[lettersize,journal]{IEEEtran}
\usepackage{amsmath,amsfonts}
\usepackage{algorithmic}
\usepackage{algorithm}
\usepackage{array}
\usepackage[caption=false,font=normalsize,labelfont=sf,textfont=sf]{subfig}
\usepackage{textcomp}
\usepackage{stfloats}
\usepackage{url}
\usepackage{verbatim}
\usepackage{graphicx}
\usepackage{cite}
\usepackage[english]{babel}

\usepackage{amsmath}
\usepackage{graphicx}
\usepackage{multirow}
\usepackage[colorlinks=true, allcolors=blue]{hyperref}

\usepackage{array}
\newcolumntype{L}[1]{>{\raggedright\let\newline\\\arraybackslash\hspace{0pt}}m{#1}}
\newcolumntype{C}[1]{>{\centering\let\newline\\\arraybackslash\hspace{0pt}}m{#1}}
\newcolumntype{R}[1]{>{\raggedleft\let\newline\\\arraybackslash\hspace{0pt}}m{#1}}

\begin{document}


\title{Choose Your Weapon: \\Survival Strategies for Depressed AI Academics}
\author{
    \IEEEauthorblockN{Julian Togelius, \emph{Senior Member, IEEE}}, 
        \IEEEauthorblockN{Georgios N. Yannakakis, \emph{Fellow, IEEE}\thanks{JT is with New York University, GNY is with University of Malta. The word ``depressed'' in the title does not refer to either the clinical or economic concept of depression, but rather the word's everyday use as signifying an unhappy and/or hopeless state of mind.}} 
    \\
}

\maketitle



\section{Introduction}\label{sec:introduction}

\IEEEPARstart{A}{s} someone who does Artificial Intelligence (AI) research in a university, you develop a complicated relationship to the corporate AI research powerhouses, such as Googe DeepMind, OpenAI, and Meta AI. Whenever you see one of these papers that train some kind of gigantic neural net model to do something you were not even sure a neural network could do, unquestionably pushing the state of the art and reconfiguring your ideas of what is possible, you get conflicting emotions. On the one hand: it is very impressive. Good on you for pushing AI forward. On the other hand: how could we possibly keep up? As an AI academic, leading a lab with a few PhD students and (if you're lucky) some postdoctoral fellows, perhaps with a few dozen Graphics Processing Units (GPUs) in your lab, this kind of research is simply not possible to do.

To be clear, this was not always the case. As recently as ten years ago, if you had a decent desktop computer and an internet connection you had everything you needed to compete with the best of researchers out there. Ground-breaking papers were often written by one or two people who ran all the experiments on their regular workstations. It is useful to point this out particularly for those who have come into the research field within the last decade, and for which the need for gigantic compute resources is a given.

If we have learned one thing from deep learning \cite{lecun2015deep}, it is that scaling works. From the ImageNet~\cite{russakovsky2015imagenet} competitions and their various winners to ChatGPT, Gato~\cite{reed2022generalist}, and most recently to GPT-4~\cite{openai2023gpt4}, we have seen that more data and more compute yield quantitatively and often even qualitatively better results. (By the time you are reading this, that list of very recent AI milestones might very well be outdated.). Of course there are improvements to learning algorithms and network architectures as well, but these improvements are mostly useful in the context of the massive scale of experiments. (Sutton talks about the ``Bitter Pill'', referring to the insight that simple methods that scale well always win the day when more compute becomes available~\cite{sutton2019bitter}.) A scale that is not achievable by academic researchers nowadays. As far as we can tell, the gap between the amount of compute available to ordinary researchers and the amount available to stay competitive is growing every year.

This goes a long way to explain the resentment that many AI researchers in academia feel towards these companies. Healthy competition from your peers is one thing, but competition from someone that has so much resources that they can easily do things you could never do, no matter how good your ideas are, is another thing. When you have been working on a research topic for a while and, say, DeepMind or OpenAI decides to work on the same thing, you will likely feel the same way as the owner of a small-town general store feels when Walmart sets up shop next door. Which is sad, because we want to believe in research as an open and collaborative endeavor where everybody gets their contribution recognized, don't we?

So, if you are but a Professor, with a limited team size and limited compute resources, what can you do to stay relevant in face of the onslaught of incredibly well-funded research companies? This is a question that has been troubling us and many of our colleagues for years now. Recent events, with models such as GPT-4 being shockingly capable and shockingly closed-sourced and devoid of published details, has made the question even more urgent. We have heard from multiple researchers at various levels of seniority, both in-person and via social media, who worry about the prospects of doing meaningful research given the lack of resources and the unfair competition from big tech companies.

Let us make this clear at the outset: both of us are secure. We hold tenured academic Professorships and we rose up on the academic ladder pretty fast, in part because of finding an academic niche: we systematically pushed the envelope of AI in the domain of video games. While we obviously care about continuing to do relevant AI research ourselves, we are writing this mostly for our more junior colleagues, postdocs and doctoral students, who may wonder about which career path to choose. Is it worthwhile to go into academia, or is it better to join a big tech company, or maybe kick off a startup? Is a career in AI a good idea, or is it better to become a plumber? Should you be a cog in the machinery, or a rebel? (It's usually easier to be a rebel when you have nothing to lose, which is either at the beginning of your career or when you have tenure.) As skilled as one may be, is this glorious battle to stay competitive lost already? Are we about to lie here, obedient to our laws? This Point of View article is partly meant as serious advice, and partly as emotional encouragement, but perhaps most of all to start a discussion with all of you so we improve our position as academics before the battle is long lost. We do not wish to stop the evolution of AI technology (even if we could); quite the contrary: we wish to discuss the strategies that will equip as many as possible to be part of this journey. While the challenges are real and many, we both feel the are even more opportunities and the time is right to grab them! 

In the remainder of this article we list a number of ideas (or strategies) for what to do if you are an AI academic despairing about your options. These options are presented in no particular order. We also don't make any particular recommendations here or ranking the options for you. It is up to you to pick one, more than one, or none of them as your favourite direction. Towards the end of the article, however, we discuss what big tech companies and universities can do to help the situation. There, we make some specific suggestions. 

\section{Give Up!}

Giving up is always an option. Not giving up on doing research, but giving up on doing things that are really impactful and pushing the envelope. There are still plenty of technical details and sub-sub-questions to publish papers about in mid-tier journals and conferences. Please note, however: (1) This works best if you already have a secure permanent position and you do not care much about promotions, (2) this wasn't really what you dreamed of doing when you decided on a research career, right? Forcing yourself to reframe your research agenda because of this fierce competition is similar to adjusting your research to the priorities of funding bodies like the European Commission or US National Science Foundation. At least going for the latter might secure some funding for your lab which can, in turn, help you work with some talented AI researchers and doctoral students. It is important to note that we both consider ourselves lucky enough as we have coordinated or have been part of several small- and large-scale research projects\footnote{Examples include the H2020 AI4Media (\url{https://www.ai4media.eu/}) and the FP7 C2Learn (\url{http://project.c2learn.eu/}) projects.} that allowed us to support our research agendas and helped us (in part) to secure our positions.  

\section{Try Scaling Anyway}

Going head-to-head with an overwhelming competition is an admirable sentiment. If scaling works, let's do it in our university labs! Let's go tilting at windmills (GPU fans)!

The most obvious problem is access to  central processing units (CPUs) and GPUs. So, let's say you secure \$50k of funding for cloud compute from somewhere and go ahead running your big experiment. But this is a very small amount of money compared to what training something like GPT-3 costs. The recent open AI agent that learned to craft a diamond pickaxe in Minecraft required training of 9 days on 720 V100 GPUs~\cite{baker2022video}; this amounts to a few hundred thousand dollars for a single experiment. Not even prestigious European Research Council (EU) or National Science Foundation (US) grants can support such a level of investment. Still, spending \$50k on cloud compute will give you significantly more compute than a bunch of gaming PCs taped together, so you could scale at least a little bit. At least for that very experiment. But as we all know, most experiments don't work the first time you try them. For every big successful experiment we see reported, we have unreported months or maybe years of prototypes, proofs of concept, debugging, parameter tuning, and failed starts. You need this level of compute available constantly.

The less obvious problem is that you need the right kind of team to build experimental software that scales, and that is generally not compatible with academic career structures. Most of the members of a typical academic research lab in computer science are PhD students that need to graduate within a few years, and need to have an individual project to work on which results in multiple first-author papers so they can get a job afterwards. A large-scale AI project typically means that most members of the team work for many months or years on the same project, where only one of them can be the first author on the paper. The team will probably also include people who do ``mundane'' software engineering tasks that are crucial to the success of the project, but which are not seen as AI research in themselves. The structures needed for successful large scale projects are simply not compatible with the structures of academia.

\section{Scale Down}

One popular way to bypass the issue is to focus on simple yet representative (toy) problems that will either prove the benefits of a new approach theoretically or showcase the comparative advantages of a novel method. Indicatively, a recent paper on Behaviour Transformers \cite{shafiullah2022behavior} showcased the benefits of the method on a toy navigation task that only took a simple multi layer perceptron to solve. A similar approach was later used in \cite{pearce2023imitating}. Both studies will likely be impactful despite the limited scale because they demonstrated the capacity of the algorithms in popular game and robotic benchmark problems that require large models and significant compute to train. In \cite{paster2022you} we observe the same pattern once again: a case is made in a toy (gambling) environment but the impact, one would argue, comes from the comparative advantages the algorithm shows in more complex but computationally heavy problems.

A downside with this approach is that people are wowed by pretty colors in high resolution, and take a real car navigating a road more seriously than a toy car, even though the challenges may be the same. So you will get less media exposure, perhaps less funding. There are also domains, such as language, which are very hard to scale down beyond some limit.

\section{Reuse and Remaster}

A key reason that AI has advanced so rapidly over the last decade is that researchers make their code and models available to the scientific community. Model sharing and code accessibility was neither the norm nor the priority of AI researchers back in the days. Having access to pretrained large models like ViT in vision \cite{dosovitskiy2020image} or the Llama family for text~\cite{touvron2023llama} saves you time and effort as you can simply resue them, and fine-tune them for your own specific problem. Arguably, one needs to assume that the representations of those large models is general enough to be able to perform well to your downstream task with limited training. Unfortunately the fine-tuning and post-hoc analysis of a large model is sometimes not enough for good performance, especially if your domain is quite different from what they were pre-trained for. Relying on pre-trained models is therefore limiting the scope of research you can do.

\section{Analysis Instead of Synthesis}

Another thing one can do with the publicly available pretrained models is to analyze them. While this may not directly contribute to new capabilities, it can still make scientific progress. The current state of things is that we have great models for text and image generation publicly available, but we don't understand them very well. You could even argue that we barely understand them all. Let's face it: a transformer is not an intuitive thing to anyone, and the scale of data these models are trained on is almost incomprehensible in itself. There is plenty of work to do in analyzing them, for example by probing them in creative ways, and developing visualizations and conceptual machinery to help us understand them.

One can do analysis with different mindsets. Trying to find and describe specific circuits and mechanisms that have been learnt is useful, and can help us (well, someone else, with resources) to create better models in the future. But one can also play the role of the gadfly, incessantly finding ways to break them! This is scientifically and societally valuable, no matter what those who try to make a business out of large models say. But it might not be the kind of research you want to do.

\section{RL! No Data!}

One might scale down one's requirements with respect to data and instead approach AI problems through the lens of (online) reinforcement learning (RL). Following the RL path might allow you to bypass issues related to data availability, analysis, storage and handling; it does not however minimize the computational effort required necessarily. In fact, even the most efficient RL methods are known to be computationally heavy as the very process of exploration is costly. Moreover, shaping a reward function often involves forms of black art (informally) or practical wisdom (more formally). That is, a researcher often needs to continuously run lengthy experiments with different types of reward (among other hyperparameters) for a breakthrough result. So ultimately one has to downscale the complexity of the problem once again. The bottom line is that if you want to break free from large data sets you might be still faced with large compute requirements unless you work on simple (toy) problems, specialized domains, or work with small models; the next section is dedicated to the latter strategy.

\section{Small Models! No Compute!}

Another valid strategy is to compromise on model scale to save on compute. There are many circumstances where you want or need a smaller model. Think of the smallest possible models that are capable of solving a problem or completing a task. This is particularly important to and relevant for real-world applications. \emph{In-the-wild} domains such as games, internet of things, and autonomous vehicles could allow AI to be deployed next to their end user and the data the user generates, i.e. at the edge of the network. This is often called \emph{edge AI} \cite{li2019edge}, the operation of AI applications in devices of the physical world is possible when memory requirements are low and inference occurs rapidly. Neuroevolution and neural architecture search \cite{li2019edge}, and knowledge distillation \cite{gou2021knowledge,makantasis2021privileged} methods are only a few of the available methods for edge AI. Note that beyond learning more from smaller models one could also attempt to learn more from less data \cite{kaci2011working}.   
Following this research path may lead to significant into models' inner workings. Studying small AI models makes the analysis far easier and increases the explainability of whatever the model does. Moreover, deploying such models on devices helps with privacy concerns. You can also argue for small models from the perspective of \emph{green AI}~\cite{schwartz2020green}~, as it minimizes the environmental footprint of the research. Obviously there are limits to what a small model is capable of doing but the importance of this research direction, we feel, will be growing drastically over the years.   

\section{Work on Specialized Application Areas or Domains}

One rather efficient strategy is to pick a niche but somewhat established area of research–--that is likely beyond the immediate interest of the industry---and try to innovate within and through that area. It is often a successful strategy to bring and test your ideas to an entirely new domain but it is less often that the outcomes will have a large impact beyond that domain. There are plenty of examples of niche areas eventually becoming dominant due to the push of a few dedicated researchers. We are both currently mostly taking this strategy: we have the AI for games community as primary scientific community where we can perform state-of-art work, as few large companies put serious efforts into modern AI for games.

Think of video games as a domain that penetrated the research communities of robotics and computer vision back in early 00s, and again with video games as deep RL benchmarks after 2015. Think of neural networks and deep learning methods that came to dominate communities invested in support vector machines and regression models (e.g. NeurIPS a decade ago). Also think of the ways reinforcement learning and deep learning have altered the core principles of multi-agent learning and cognitive/affect modeling in communities represented by the AAMAS, ACII and IVA conferences, for instance.

A core downside to this strategy is the difficulty getting your paper accepted in the kind of large venues that are most influential in AI, such as NeurIPS, AAAI, ICML and IJCAI. Your paper and its results might end up sitting out-of-the-interest-distribution. It is, however, very possible to start your own community with its own publication venues.

If you do not have the requisite domain expertise---and/or datasets---yourself, you can fruitfully approach domain experts to collaborate. The good news is that as an academic, you have plenty of such experts in other departments of your university or institute and they all have interesting AI problems to solve if you spend some time talking to them. One of the authors recently ran in to an anthropologist and an analytical chemist in a corridor, and started discussing projects that would include all three. Another example is a recent collaboration of one of the authors with urban designers resulting in the reconstruction of urban areas around MIT and Harvard for improving the comfort levels of Bostonians \cite{galanos2021arch}. 

These projects may not end up advancing the state of AI much, but may make big differences in the particular disciplines. And sometimes big AI advances come from application-specific work.

\section{Solve Problems Few Care About (For Now!)} 

While focusing on an established niche or application field is a relatively safe strategy, a somewhat riskier one is to find a niche or application that does not exist yet. Basically, focus on a problem that almost no-one sees the importance of, or a method that nobody finds promising. 

One approach is to go looking for applications that people have not seriously applied AI to. A good idea is to look into a field that is neither timely nor ``sexy''. The bet here is that this particular application domain will become important in the future, either in its own right or because it enables something else. We both took this path. Procedural content generation for games was a very niche topic 15 years ago and we helped nuild a research community around it\cite{togelius2011search,yannakakis2011experience}; recently it has become more important not only for the games industry, but also as a way to help generalize (deep) reinforcement learning \cite{risi2020increasing,team2023human}. Research on reinforcement learning is a core AI topic with thousands of papers published per year, lending more importance to this once somewhat obscure topic. This high-risk high-gain mindset might lead to a lonely path that nevertheless could end up being highly rewarding in the long run.

So, look around you, and talk to people who are not AI researchers. What problem domains do you see where AI is rarely applied, and which AI researchers seem to not know or care about? Might someone care about these domains in the future? If so, you may want to dig deeper in one of those domains.

\section{Try Things that Shouldn't Work}

Another comparative advantage of small academic teams is the ability to try things that ``shouldn't work'', in the sense that they are unsupported by theory or experimental evidence. The dynamics of large industry research labs are typically such that researchers are incentivized to try things that are likely to work; if not, money is lost. In academia, failure can be as instructive and valuable as success and the stakes are lower overall. Many important inventions and ideas in AI come from trying the ``wrong'' thing. In particular, all of deep learning stems from researchers stubbornly working on neural networks even though there were good theoretical reasons why they shouldn't work.

\section{Do Things that Have Bad Optics} 

The larger and more important a company is, the more constrained it is by ethics and optics. Any company is ultimately responsible to their shareholders, and if the shareholders perceive that the company suffers ``reputational damage'' they can easily fire the CEO. So large companies will try to avoid to do anything that looks bad. To get around this, large companies sometimes fund startups to do their more experimental work that might go wrong (think Microsoft and OpenAI). But even such plays have limits, as bad PR can come washing back like the tide in San Francisco Bay.

As an individual researcher, who either has no position or who already has a secure position, you have nothing to lose. You can do things that are as crazy as you like. You are only constrained by the law and your own personality. Now, we are in no way arguing that you should do research that is unethical. By all means, try to do the right thing. But what you find objectionable might be very different from what a group of mostly-white liberal overeducated engineers in coastal USA find objectionable. The PR departments, ethics committees, and boards of directors of the rich tech companies espouse a very particular set of values. But the world is large, and full of very different people and cultures. So there is a big opportunity to do research that these tech companies will not do even though they could.

As an example of a project that exploits such an opportunity, one of us participated in a project critically examining the normativity of the ``neutral English'' in current writing support systems by creating an autocomplete system with a language model that assumes you write in the tone of Chuck Tingle, the famous author of absurd sci-fi political satire gay erotica~\cite{khalifa2017deeptingle}. Our guess is that this project would not have been cleared for publication by Amazon or Google. Another example is this very paper.

Similarly, you may find that you deviate from the cultural consensus in big tech companies regarding topics relating to nudity, sexuality, rudeness, religion, capitalism, communism, law and order, justice, equality, welfare, representation, history, reproduction, violence, or something else. As all AI research happens in and is influenced by a cultural and political context, see your deviation from the norm as an opportunity. If you can't do the research they couldn't do, do the research they wouldn't do.

\section{Start it Up; Spin it Out!}

By now it should be rather clear that academia is somewhat, paradoxically, limiting academic AI research. Even if one manages to secure large-scale multimillion projects this covers only a fraction of human and computational resources that are necessary for contemporary AI research, and the career structures and IP rights regimes of universities often impose further limits. One popular alternative among AI scientists is to spin out their idea from their university lab and found a company that will gradually transfer AI research to a set of commercial-standard services or products. Both authors have been part of this journey through co-founding modl.ai \cite{pedersen2022experience} and have learned a lot from this.

Being part of the applied AI world offers many benefits. In principle you get access to rich data from real-world applications that you wouldn't be able to have otherwise. Moreover your AI algorithms are tested on challenging commercial-standard, applications and have to be operational in the wild. Finally, you usually gain access to more compute and, if the start-up scales up, growing access to human resources.

This journey is far from straightforward, however, as there are several limiting factors to consider. First, not all research ideas are directly applicable to a startup business model. Your best research ideas might be brilliant in terms of understanding the world, or at least getting published in highly prestigious venues, but that does not mean that one can easily make products out of them. Second, many outstanding results one obtained in the lab today may have to go through a long runway until they turn into a business case of some sort. Most startups do development rather than research, as the runways are short and you need to have a functioning product, preferably with some market traction, before the next funding round in two years or so. Third, even if you do get some investment, this does not mean you have an unlimited compute budget. With seed grants often in the range of a few millions, this does not buy you the capacity to do OpenAI-level experiments, especially as you need to pay real salaries (not PhD stipends) to your employees. Fourth, not every AI academic enjoys this type of an adventure. At the end of the day most academics have long agreed on their priorities when they opted to follow the academic career path. You don't become a professor for the money. The security of an academic environment (given that it is both safe and creative), means to some far more than any potentially higher salary or other corporate benefits.

Here, we might point out that both of us publish many more papers with our academic research teams than with the company we co-founded and work part-time at. On the other hand, we believe we have more direct impact on the games industry through our company.

\section{Collaborate, or Jump Ship!} \label{sec:collaborate}

If none of the above options work for you and you still want to innovate though large scale methods that are trained on lots of data you can always collaborate with those that have them both: compute and data. There are several ways to move forward with this approach. 

Universities in the vicinity of leading AI companies have a comparative advantage as local social networks and in-person meetings make the collaboration easier. Researchers from remote universities can still establish collaborations though research visits, placements and internships as part of a joint-research project. More radically, some established AI professors decide to dedicate some (if not all) of their research time to an industrial partner or even move their entire lab in there. Results from such partnerships, placements or lab transfers can be astonishing \cite{vaswani2017attention,wang2016dueling}. At a glance, this looks like the best way forward for AI academics, however, 1) the generated IP cannot always be published and 2) not everyone can or want to work in an industry-based AI lab.

One might even argue that innovation should be driven by public institutions as supported by the industry, not the other way around. It is arguably the university’s responsibility to maintain (part of, or some of) the talented AI researchers it educates (academics and students) and the IP they generate. Otherwise AI education and research will eventually become redundant within a University environment. This would be bad for everyone, as knowledge would be less open, and there would be no-one to train the next generation of AI researchers. Next, let's look at this relationship more closely and outline ways industrial corporations and universities may be able to help.

\section{How Can Large Players in Industry Help?}

It is not clear that large companies with well-financed AI labs actually want to help alleviate this situation. Individual researchers and managers might care about the depression of academic AI research, but what the companies care about is the bottom line and shareholder value, and having a competitive academic research community might or might not be in their best interest. However, to the extent that large private sector actors do care, there are multiple things they can do.

At the most basic level, open-sourcing models, including both weights and training scripts, helps a lot. It allows academic AI researchers to study the trained models, fine-tune them, and build systems around them. It still leaves academic researchers uncompetitive when it comes to training new models, but it is a start. To their credit, several large industrial research organizations regularly release their most capable models publicly; Meta in particular stands out. Others don't, and could rightly be shamed for not doing so. In particular if their name implies some degree of openness.

The next step for remedying this situation is to collaborate with academia. As discussed earlier (see Section \ref{sec:collaborate}) some large institutions regularly do this, mostly through accepting current PhD students as interns, allowing these students to do large-scale work. Some offer joint appointments to certain academic researchers, and a few even occasionally offer research grants. All of this is good, but more can be done. In particular there could be mechanisms where academics initiate collaborations by proposing work they would do collaboratively, and there could be more stable research funding mechanisms.

Going even further, private companies that really wanted to help mend this academia-industry divide could choose to work in public: post their plans, commit code, models, and development updates to public repositories and allowing academics to contribute freely. This is not how most companies work, and often they have good reasons for their secrecy. On the other hand, a lot could be gained from having academics contributing to your code and training for free.

\section{How Can Universities Help?}

As much as industry might be willing to help, the primary initiative should come from those universities that wish to drive innovation. Universities have a strong initiative to stay on top of (or if possible be in the driving seat for) AI research for many reasons, including their role in educating students who will look for jobs in a world transformed by AI, and the many ways in which AI systems transform education~\cite{lim2023generative}. It is worth noting that many of the most influential papers in AI involve a university department. Those papers are typically co-authored by researchers that either collaborate with or are involved in a company. The successful examples are out there \cite{vaswani2017attention,wang2016dueling,chen2021decision}, but more is needed from the university’s end to enable such partnerships. And actually, there are many ways an academic institution can initiate and foster collaboration with the industry.  

Universities can also help their faculty manage the changed competitive landscape by encouraging and allowing them to be more risk-taking. The comparative advantage of academic researchers in AI is to do more high-risk exploration, and incentive structures at universities must change to account for this. For example, it is unreasonable to expect a steady stream of papers at top-tier conferences such as NeurIPS and AAAI; large, well-funded industry research labs will have large advantages at writing such papers. Similarly, the grant funding structure is such that it rewards safe and incremental research on popular topics; this seems to be an inherent feature of the way grant applications are evaluated, and it is unlikely to change however often funding agencies use words like ``disruptive''. The kind of research that is favored by some of the most traditional (closed-call) grant mechanisms is mostly the kind of research where academic AI researchers will not be able to compete with industry. Therefore, universities should probably avoid making grant funding a condition for hires and promotions. If universities are serious about incentivizing their faculty to leverage their competitive advantage, they should reward trying and failing and promote high-risk high-gain funding schemes and research initiatives. It is then likely that funding agencies will follow the trend and invest even more on basic and blue sky research. 

Such a mindset might further open the possibilities for academics to attract large amounts of funding and collectively start building their own large (foundation) models that would be entirely open to any researcher. European research funding, for instance, has long supported the AI-on-Demand Platform\footnote{\url{https://www.ai4europe.eu/}}---a community-driven channel featuring open access AI tools---that could host such collaborative efforts on model building and sharing. The seeds of collaborative open-source projects are already planted; think of StarCoder, the recent large model built by an open-science community involving both universities and industrial partners \cite{li2023starcoder}. We feel it is only a matter of time that more and larger academic-driven models and data will be shared openly. 

\section{Parting Words}

We wrote this Point of View article with several purposes in mind. First, to share our concerns with other fellow AI researchers with a hope of finding a common cause (and a collective remedy?) as a community. Second, to offer a set of guidelines based on our own experiences but also the discussions we had in the academic and industrial AI venues we participate or organize. Third, to spur an open dialogue and solicit ideas for potentially more efficient strategies for us all. Arguably, the list of strategies we ended up discussing here are far from inclusive of all possibilities that are available out there; we believe, however, that they are seeds of a conversation that---in our opinion---is very timely. 

\section{Acknowledgements}

We would like to thank all anonymous and eponymous reviewers for their insightful comments. This work has been supported by NSF Award 1956200 and by a GoodAI award (JT) and from the European Union’s Horizon 2020 programme under grant agreement No 951911 (GNY). 

\bibliographystyle{plain}
\bibliography{sample}

\begin{IEEEbiography}[{\includegraphics[width=1in,height=1.25in,clip,keepaspectratio]{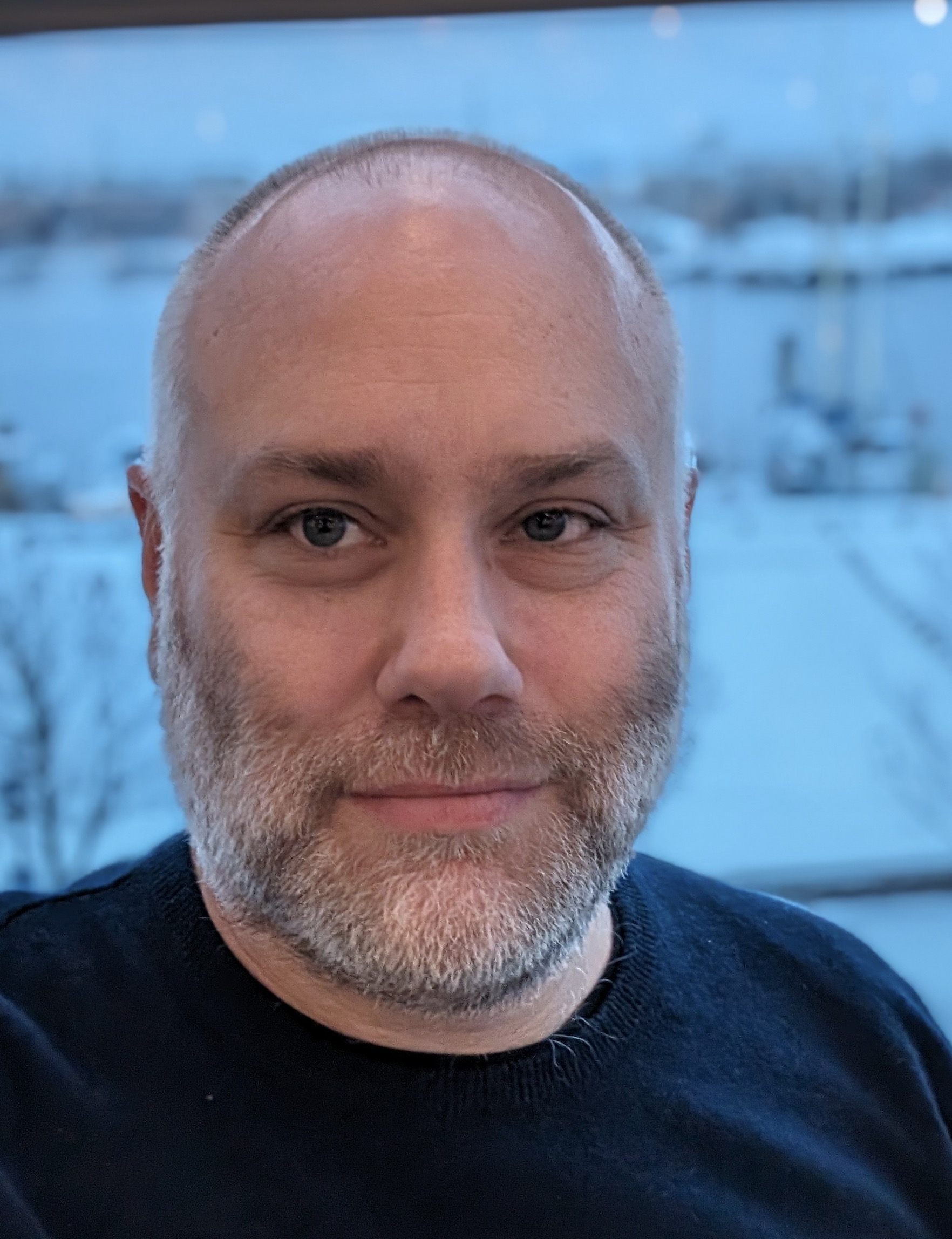}}]
{Julian Togelius} (S'05-M'07-SM'22) is an Associate Professor in the Department of Computer Science and Engineering, New York University, and a co-founder of modl.ai. He works on artificial intelligence for games and on games for artificial intelligence. His current main research directions involve procedural content generation in games, general video game playing, player modeling, and fair and relevant benchmarking of AI through game-based competitions. Additionally, he works on topics in evolutionary computation, quality-diversity algorithms, and reinforcement learning. From 2018 to 2021, he was the Editor-in-Chief of the IEEE Transactions on Games. Togelius holds a BA from Lund University, an MSc from the University of Sussex, and a PhD from the University of Essex. He has previously worked at IDSIA in Lugano and at the IT University of Copenhagen.

\end{IEEEbiography}

\begin{IEEEbiography}[{\includegraphics[width=1in,height=1.25in,clip,keepaspectratio]{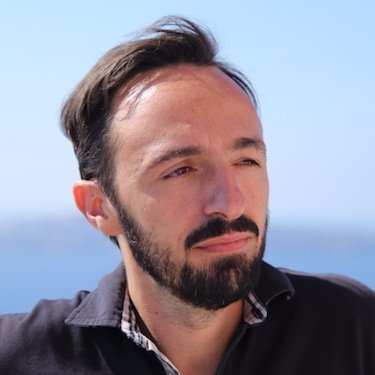}}]{Georgios N. Yannakakis}(S'04-M'05-SM'14-F'24) is a Professor at the Institute of Digital Games, University of Malta, and a co-founder of modl.ai. He does research at the crossroads of artificial intelligence, computational creativity, affective computing, advanced game technology, and human-computer interaction. He has published more than 350 papers in the aforementioned fields and his work has been cited broadly. He is currently the Editor in Chief of {\sc IEEE Transactions on Games} and an Associate Editor of {\sc IEEE Transactions on Evolutionary Computation}. 
\end{IEEEbiography}

\end{document}